\documentclass[12pt]{article}
\usepackage{amsfonts}
\usepackage{amsmath}
\usepackage{amssymb}
\usepackage{latexsym}
\usepackage[dvips]{graphicx}
\usepackage{color}

\usepackage{soul}

\newcommand{\hpi}{\hat{\Pi}}

\newcommand{\polv}{{\phi}}
\newcommand{\hpolv}{\hat{\phi}}

\newcommand{\Lpar}{{\tilde L}}
\newcommand{\LparP}{{{\tilde L}_P}}
\newcommand{\Lpareff}{{{\tilde L}_{\text{eff}}}}

\newcommand{\be}{\begin{equation}}
\newcommand{\ee}{\end{equation}}
\newcommand{\bea}{\begin{eqnarray}}
\newcommand{\eea}{\end{eqnarray}}

\newcommand{\BbbR}{\mathbb{R}}
\newcommand{\BbbZ}{\mathbb{Z}}

\def\gsim{\raise0.3ex\hbox{$>$\kern-0.75em\raise-1.1ex\hbox{$\sim$}}}
\def\lsim{\raise0.3ex\hbox{$<$\kern-0.75em\raise-1.1ex\hbox{$\sim$}}}

\newcommand{\lscale}{{\bar{\mu}}}

\numberwithin{equation}{section}

\begin{document}

\begin{titlepage}
\begin{center}
{\Large {\bf Boundary conditions in quantum mechanics\\[1ex] 
on the discretized half-line}}
\vspace{1cm} 


\renewcommand{\baselinestretch}{1}
{\bf
Gabor Kunstatter${}^\dagger$
and 
Jorma Louko${}^\sharp$
\\}
\vspace*{0.7cm}
{\sl
${}^\dagger$ 
Department of Physics and
Winnipeg Institute of Theoretical Physics,\\
The University of Winnipeg,\\
515 Portage Avenue,
Winnipeg, Manitoba, Canada R3B 2E9\\
{[e-mail: g.kunstatter@uwinnipeg.ca]}\\[5pt]
}
{\sl
${}^\sharp$ 
School of Mathematical Sciences,
University of Nottingham,\\
Nottingham NG7 2RD, United Kingdom\\
{[e-mail: jorma.louko@nottingham.ac.uk]}\\ [5pt]
}
\vspace{2ex}
{\bf Abstract}
\end{center}
We investigate nonrelativistic quantum mechanics on the discretized
half-line, constructing a one-parameter family of Hamiltonians that
are analogous to the Robin family of boundary conditions in continuum
half-line quantum mechanics. For classically singular Hamiltonians,
the construction provides a singularity avoidance mechanism that has
qualitative similarities with singularity avoidance encountered in
loop quantum gravity. Applications include the free particle, the
attractive Coulomb potential, the scale invariant potential and a
black hole described in terms of the Einstein-Rosen wormhole
throat. The spectrum is analyzed by analytic and
numerical techniques. In the continuum limit, the full Robin family
of boundary conditions can be recovered via a suitable fine-tuning but
the Dirichlet-type boundary condition emerges as generic.
\vfill 
\noindent 
Published in J. Phys.\ A: Math.\ Theor.\ \textbf{45}, 305302 (2012)
\newline
\noindent 
{\tt http://stacks.iop.org/1751-8121/45/305302}
\newline
\noindent
January 2012; revised June 2012 
\end{titlepage}

\section{Introduction}
\label{sec:intro}

It is often useful and sometimes necessary to study quantum mechanics
on the half-line $[0,\infty)$. The boundary, which 
has been taken to be at the origin without loss of generality, 
can represent the location of a singularity in the
potential or an infinite potential barrier (impenetrable wall) in the
presence of an otherwise non-singular potential. The form of the
Hamiltonian determines what conditions on the wave functions at the
boundary lead to a quantization with unitary time evolution, if indeed
such quantizations exist.  Textbook accounts can be found in
\cite{reed-simonII,blabk} and accessible reviews in~\cite{extensions}.

If the Hamiltonian is essentially self-adjoint, no boundary conditions
are required. It may however happen that the Hamiltonian has a family
of self-adjoint extensions, each specified by a boundary condition. As
an example, consider a particle in a smooth potential.  If the
potential is sufficiently well-behaved at $x\to\infty$ so that no
boundary conditions are needed there, the self-adjoint extensions are
parametrized by a parameter 
$L \in \BbbR \cup \{\infty\} \simeq S^1 \simeq \text{U}(1)$, 
such that the corresponding boundary condition on the wave function is
the Robin boundary condition, 
\begin{align}
\psi(0)+L\psi'(0) =0 \ .
\label{Robin1}
\end{align}
The special case $L=0$ gives the Dirichlet boundary condition,
$\psi(0)=0$, and the special case $L=\infty$ gives the Neumann
boundary condition, $\psi'(0)=0$. All values of $L$ make the
time evolution unitary, and $L$ may be physically interpreted in terms
of the phase shift of a wave packet on its reflection at the origin.

From a purely mathematical point of view all the self-adjoint
extensions are created equal, none are preferred. In most physical
situations, however, Dirichlet boundary conditions seem to be
considered `natural'. One argument to this end comes from regarding a
singular potential or an infinite wall as a mathematical idealization of a
physical situation in which the potential 
is regular~\cite{walton}.  Assuming that the Hamiltonian for the
regular potential is already essentially self-adjoint and requires no
boundary conditions, the `physical' boundary conditions in the
singular potential approximation may then be defined as those obtained
when the regular potential approaches the singular one. It is a
somewhat surprising mathematical fact that recovering in this limit
anything other than the Dirichlet boundary conditions tends to require
severe fine-tuning in the way in which the limit is
taken~\cite{walton}.

The context of the foregoing discussion was conventional Schr\"odinger
quantization. In this paper we investigate a similar issue of
boundary conditions in polymer quantization, in which the continuous
variable $x \in \BbbR_+ \cup \{0\}$ is replaced by a discrete variable
and the inner product becomes a discrete sum rather than an
integral~\cite{afw,halvorson}. 
This quantization scheme
arises most naturally in loop quantum
gravity~\cite{Thiemann07,Rovelli07}. 
The term ``polymer" stems from that fact 
that the quantum states of the so-called ``polymer particle
represention'' are mathematically analoguous 
to the polymer-like excitations of quantum geometry~\cite{afw}. 
Although we retain the loop quantum
gravity motivated name, we stress that polymer 
quantization is a viable quantization
procedure in its own right, unitarily inequivalent to Schr\"odinger
quantization. It has been applied to a variety of systems, including
the harmonic oscillator~\cite{afw}, the Coulomb
potential~\cite{coulomb}, the scale invariant potential~\cite{scale invariant}, 
and the Schwarzschild spacetime as described in terms of
the dynamics of the Einstein-Rosen wormhole
throat~\cite{throata,throatb}. In the last three of these systems the
configuration space is the half-line, rather than the full real line,
and the analyses in \cite{coulomb,scale invariant,throata,throatb}
treat the restriction to a half-line by first polymer quantizing on
the full real line and then considering separately the even and odd
sectors, mimicking respectively the Neumann and Dirichlet boundary
conditions in continuum. In the case of the even sector, a potential
that is singular at the origin needs in addition to be regulated, and
this was done by the Thiemann trick~\cite{Thiemann}. Most recently, Dirichlet
boundary conditions were applied in the analysis of the statistical thermodynamics
of a polymer particle in a box~\cite{ChaconAcosta:2011vv}. 

In this paper we construct a family of polymer
Hamiltonians that is analogous to the Robin family of boundary
conditions in Schr\"odinger quantization, and we analyze the continuum
limit of these polymer theories. The construction does not rely on
extending the polymer dynamics from half-line to full line, and it
applies to potentials that are singular at the origin without a need
to introduce a separate regularization for the potential.  We shall
show that the continuum limit of our family of polymer Hamiltonians
does reproduce the Robin family of boundary conditions for the
Schr\"odinger Hamiltonian.  We shall further find that a fine-tuning
of the polymer Hamiltonian in this limit is required in order to
recover Schr\"odinger boundary conditions other than Dirichlet.  The
continuum Dirichlet boundary conditions hence emerge as generic when
approached from polymer quantization, just as they emerge as generic
when approached via regular potentials within Schr\"odinger
quantization.

As concrete examples, we consider the following one-dimensional
systems, each of which is motivated by specific physical applications:
\begin{enumerate}
\item 
The free particle in the presence of an infinitely
high potential wall. This is the prototype for any discussion of quantum
mechanics on the half line. 
\item 
The attractive Coulomb potential. This describes the spherically
symmetric sector of the hydrogen atom.  
\item 
The scale invariant potential. This has applications in physics
of polymers \cite{polymer} and in molecular
physics~\cite{molecular}, and it also arises in the context of
quantizing spherically symmetric black holes~\cite{black-hole}.
\item 
The Hamiltonian that describes the dynamics of the Einstein-Rosen
wormhole throat~\cite{throata,throatb}. 
\end{enumerate}

The paper is organized as follows: In Section \ref{sec:polhalfline} we
present the general construction of our half-line polymer Hamiltonians
for systems whose classical Hamiltonian is of the conventional
nonrelativistic form, $H = p^2 + V(x)$.
Sections 
\ref{sec:freeparticle}, 
\ref{sec:hydrogen}
and 
\ref{sec:scaleinv}
apply this construction respectively to the free particle, the
attractive Coulomb potential and the scale invariant potential.
Section \ref{sec:throat} addresses the Einstein-Rosen wormhole throat
dynamics: in this system the classical kinetic term has a more
complicated form, but we show that the construction of Section
\ref{sec:polhalfline} has nevertheless a natural
generalization. Section \ref{sec:conclusions} presents a summary and
brief concluding remarks.

We use throughout dimensionless units: for the Coulomb potential the
configuration space coordinate is twice the Rydberg radial coordinate
as in~\cite{coulomb}, and in the Einstein-Rosen throat theory the
units are the Planck units as in~\cite{throata,throatb}. Complex
conjugation is denoted by overline.

\section{Polymer half-line quantization for nonrelativistic Hamiltonians}
\label{sec:polhalfline}

In this section we present our family of half-line polymer
Hamiltonians for systems whose classical Hamiltonian consists of
kinetic and potential terms of nonrelativistic standard form.
Subsection \ref{subsec:realline} recalls the key features of polymer
quantization of this system on the full real line~\cite{afw}.  The
half-line version is given in subsection~\ref{subsec:halfline}.

\subsection{Polymer real line}
\label{subsec:realline}

We consider a system whose classical phase space is 
$\BbbR^2 = \left\{ (x,p)\right\}$ with the Poisson bracket 
$\{ x,p\}=1$ and the Hamiltonian 
\begin{align}
H = p^2 + V(x) 
\ . 
\label{eq:class-Ham}
\end{align}
We assume the potential $V$ to be defined for all $x$ and real-valued. 

The polymer Hilbert space is spanned by the basis states
\begin{align}
\psi_{x_0}(x) = \left\{
	\begin{array}{ll}
	1, & x=x_0\\
	0, & x\neq x_0
	\end{array}
\right.
\end{align}
with the inner product
\begin{align}
(\psi_x , \psi_{x^\prime}) = \delta_{x,x^\prime},
\label{eq:bohr-ip}
\end{align}
where the object on the right hand side is the Kronecker delta.
The position operator $\hat{x}$ acts by pointwise multiplication as
\begin{align}
\bigl(\hat{x} \psi\bigr)(x) := x \psi(x), 
\label{xact}
\end{align}
and the family of translation operators 
$\bigl\{\widehat{U}_\mu \mid \mu\in\BbbR\bigr\}$
is defined as in Schr\"odinger quantization by 
\begin{align}
\bigl(\widehat{U}_\mu \psi\bigr)(x) := \psi(x+\mu). 
\label{Uact}
\end{align}

In Schr\"odinger quantization the action of 
$\widehat{U}_\mu$ \eqref{Uact} is weakly continuous 
in~$\mu$, and the usual momentum operator 
$\hat{p} := -i\partial_x$ can be defined 
in terms of the translation operator by 
$\widehat{U}_\mu=e^{i\mu \hat{p}}$. 
In polymer quantization the action of $\widehat{U}_\mu$ 
is not weakly continuous in $\mu$ and a similar 
definition of a momentum operator is not available. 
Instead, we now take $\mu$ to be a fixed positive 
constant, understood as a fundamental length scale in 
the polymer quantum theory, 
and define the momentum operator and its square by 
\begin{subequations} 
\begin{align}
\hat{p} &:= \frac{1}{2i\mu}(\widehat{U}_\mu - \widehat{U}_\mu^\dagger),\\
\widehat{p^2} &:= \frac{1}{\lscale^2}(2 - \widehat{U}_{\lscale} - \widehat{U}_{\lscale}^\dagger),  
\end{align}
\end{subequations}
where $\lscale := 2\mu$. 
The polymer Hamiltonian is defined to be 
\begin{equation}
\widehat{H} := 
\widehat{T}
+ 
\widehat{V} \ , 
\label{Hpol}
\end{equation}
where $\widehat{V}$ 
acts by pointwise multiplication, 
\begin{align}
(\widehat{V} \psi)(x) := V(x) \psi(x), 
\label{Vpolact}
\end{align}
and 
\begin{align}
\widehat{T}
:= \widehat{p^2} = 
\frac{1}{\lscale^2}
(2 - \widehat{U}_\lscale - \widehat{U}_\lscale^\dagger). 
\label{Tpol}
\end{align}

$\widehat{H}$ is clearly symmetric. Its action  
decomposes the polymer Hilbert
space into a continuum of separable superselection sectors, 
where each sector consists of 
wave functions supported on the 
regular $\lscale$-spaced lattice $\left\{\Delta + m\lscale \mid m \in
\mathbb{Z}\right\}$ and the parameter 
$\Delta \in [0, \lscale)$ parametrizes the sectors. 
A~study of the deficiency indices 
\cite{reed-simonII} shows that 
$\widehat{H}$ has at least one self-adjoint extension in each superselection sector. 
In particular, the 
Kato-Rellich theorem 
\cite{reed-simonII}
can be applied as in 
\cite{scale invariant} to show that $\widehat{H}$ is essentially 
self-adjoint in every superselection sector in which the function $V$ is bounded. 

It is useful to write the Hilbert space of each superselection sector
in terms of sequences. 
For concreteness, consider the 
superselection sector $\Delta=0$. 
An orthonormal basis is given by 
$\bigl\{\psi_{m\lscale} \mid m \in
\mathbb{Z}\bigr\}$. 
Writing $\psi = \sum_m c_m \psi_{m\lscale}$, the inner product reads 
$\left( \psi^{(1)}, \psi^{(2)} \right) = 
\sum_m \overline{{c_m}^{(1)}} \, c_m^{(2)}$, and the action of 
$\widehat{H}$ reads 
\begin{align}
\widehat{H}
\, \biggl(\sum_m c_m \psi_{m\lscale} \biggr)
= \sum_m \left( \frac{2 c_m - c_{m+1} - c_{m-1}}{\lscale^2}
+ V(m\lscale) c_m \right) \psi_{m\lscale}
\ . 
\end{align}
The Hilbert space hence consists of the 
two-sided square summable sequences 
$c := {(c_m)}_{m=-\infty}^\infty$, 
and the action of $\widehat{H}$ 
reads    
\begin{equation}
\bigl(\widehat{H} c\bigr)_m 
= \frac{2 c_m - c_{m+1} - c_{m-1}}{\lscale^2}
+ V(m\lscale) c_m 
\ . 
\label{eq:fH-action2}
\end{equation}

\subsection{Polymer half-line}
\label{subsec:halfline}

We wish to modify the 
polymer Hamiltonian \eqref{eq:fH-action2} 
into a Hamiltonian 
in the Hilbert space of 
\emph{one-sided\/} square-summable sequences. 

We take the one-sided 
sequences to be of the form $c := {(c_m)}_{m=1}^\infty$ 
and the inner product to read 
$\left( d, c \right) = 
\sum_{m=1}^\infty \overline{d_m} \, c_m$. 
We define the one-parameter family of modified 
Hamiltonians $\bigl\{\widehat{H}_\alpha \mid \alpha\in\BbbR\bigr\}$ 
by 
\begin{align}
\bigl(\widehat{H}_\alpha c\bigr)_m 
:= 
\begin{cases}
{\displaystyle \frac{2 c_m - c_{m+1} - c_{m-1}}{\lscale^{2}} + V(m\lscale) c_m}
& 
\text{for $m>1$;} 
\\[2ex]
{\displaystyle \frac{(2-\alpha) c_1 - c_{2}  }{\lscale^{2}} + V(\lscale) c_1}
& 
\text{for $m=1$.} 
\end{cases} 
\label{eq:fHm-action}
\end{align}
Each $\widehat{H}_\alpha$ is symmetric, and 
it can be verified as with the two-sided sequences that 
each $\widehat{H}_\alpha$ has at least one self-adjoint extension. 
If $V=0$, it can be explicitly verified that 
there are no nonvanishing normalizable solutions to
$\widehat{H}_\alpha c = \pm i c$, 
and in this case $\widehat{H}_\alpha$ is hence essentially self-adjoint. 
It follows by the Kato-Rellich theorem \cite{reed-simonII}
that $\widehat{H}_\alpha$ is essentially self-adjoint whenever the set 
$\bigl\{V(m\lscale) \mid m=1,2,\ldots\bigr\}$ is bounded. 

The motivation for the definition of 
$\bigl(\widehat{H}_\alpha c\bigr)_m$ by \eqref{eq:fHm-action} 
is that $\bigl(\widehat{H}_\alpha c\bigr)_m$ agrees with 
$\bigl(\widehat{H}c\bigr)_m$ \eqref{eq:fH-action2}
for $m\ge1$ 
if we envisage there to be a fictitious lattice point at $m=0$ such
that $c_0 = \alpha c_1$. This means that 
$\alpha\in\BbbR$ can be regarded as a lattice counterpart of the continuum
theory self-adjoint extension parameter~$L$. The case $\alpha=0$ is
analogous to continuum Dirichlet, with the continuum boundary at
$m=0$.  The cases $\alpha=\pm1$ are respectively analogous to the continuum
Neumann and continuum Dirichlet, with the continuum boundary half-way
between $m=0$ and $m=1$.

Two comments are in order. 
First, given the definition~\eqref{eq:fHm-action}, 
the coefficient of $\overline{d_1} \, c_1$ in the 
inner product 
$\left( d, c \right) = 
\sum_{m=1}^\infty \overline{d_m} \, c_m$ 
cannot 
be changed
without losing symmetricity of~$\widehat{H}_\alpha$. 
Second, although the definition of $\widehat{H}_\alpha$ 
is motivated by a fictitious lattice point at $m=0$, 
the actual definition involves $V$ only at lattice points 
$x = m \lscale$ with $m\ge1$. 
$\widehat{H}_\alpha$~is hence well defined 
even if the classical potential is singular at $x=0$.

\section{Free particle}
\label{sec:freeparticle}

As the first example we consider the free particle, $V=0$. 
We begin by recalling relevant facts from Schr\"odinger 
quantization on the 
half-line \cite{reed-simonII,blabk,extensions} 
and from polymer quantization on the full real line~\cite{afw}. 
We then analyze $\widehat{H}_\alpha$ 
\eqref{eq:fHm-action} and its continuum limit.

\subsection{Schr\"odinger half-line}

In Schr\"odinger quantization, the 
Hamiltonian operator of a free particle on the half-line is 
$\widehat{H}_{\text{free}} := - d^2/dx^2$, acting in the Hilbert space 
$L_2(\BbbR_+, dx)$. 
$\widehat{H}_{\text{free}}$ is symmetric and has a $\text{U(1)} \simeq S^1$ of 
self-adjoint extensions, specified by the boundary condition~\eqref{Robin1}. 

The spectrum of each extension contains the positive continuum. 
The extensions with 
$0 < L < \infty$ have in addition a discrete 
ground state of energy~$-L^{-2}$. The spectral decomposition of the 
identity in terms of the eigenfunctions 
can be found in~\cite{titchmarsh-eigenI}.

\subsection{Polymer real line}

Polymer quantization on the full real line can be analyzed by
interpreting the two-sided square summable sequence $c :=
{(c_m)}_{m=-\infty}^\infty$ as the coefficients in the Fourier
expansion $\chi(\varphi) = {(2\pi)}^{-1/2} \sum_m c_m e^{im\varphi}$
of the function $\chi \in L_2(S^1)$, so that in this realization
$\bigl(\widehat{H}\chi\bigr)(\varphi) = 2\lscale^{-2} \left(1 -
\cos\varphi\right)\chi(\varphi)$. 
It is immediate that
$\widehat{H}$ is essentially self-adjoint and its spectrum is the
continuum $(0,4\lscale^{-2})$.

\subsection{Polymer half-line}

The half-line polymer Hamiltonian is given by 
$\widehat{H}_\alpha$ \eqref{eq:fHm-action} with $V=0$. 
As noted above, $\widehat{H}_\alpha$ is essentially self-adjoint.

\subsubsection{Spectrum}

We wish to find the spectrum. 

Let $E\in\BbbR$ denote the energy eigenvalue, which may be a
proper eigenvalue in the discrete spectrum or an improper
eigenvalue in the continuous spectrum. 
The eigenvalue equation reads 
\begin{align}
\widehat{H}_\alpha c = Ec \ . 
\label{eq:free-eigenvalue-eq}
\end{align}
If $E\ne0$ and $E \ne 4\lscale^{-2}$, 
the solution to \eqref{eq:free-eigenvalue-eq} as a difference
equation is 
\begin{align}
c_m = a  A_+^m + b A_-^m 
\ , 
\label{eq:free-gensol}
\end{align}
where $a$ and $b$ are constants, 
\begin{align}
A_{\pm} := \left(1-\tfrac12\lscale^2E\right) 
\pm 
\sqrt{\left(1-\tfrac12\lscale^2E\right)^2-1}
\ , 
\label{eq:Apm}
\end{align}
and evaluating \eqref{eq:free-eigenvalue-eq} at the lattice point 
$m=1$ by \eqref{eq:fHm-action} shows that 
\eqref{eq:free-gensol} must in addition satisfy 
\begin{align}
c_2 = \left(2 - \alpha - \lscale^2E \right)c_1
\ , 
\label{eq:free-m=1sol}
\end{align}
so that for each $\alpha$ there is only one linearly independent solution. 
If $E=0$ or $E = 4\lscale^{-2}$, we proceed similarly, finding that 
the only linearly independent solution 
to \eqref{eq:free-eigenvalue-eq} as a difference equation is 
\begin{equation}
c_m = 
\begin{cases}
\alpha + (1-\alpha)m  & 
\text{for $E=0$;} 
\\[1ex]
{(-1)}^m \left[\alpha - (1+\alpha)m \right]  & 
\text{for $E = 4\lscale^{-2}$.} 
\end{cases} 
\label{eq:free-cm-exceptional}
\end{equation}
The spectrum is now found by considering the normalizability 
properties of these solutions. 

Suppose first that 
$0 < E < 4\lscale^{-2}$. 
We parametrize $E$ as 
\begin{align}
\lscale^2E = 4 \sin^2(\theta/2)
\ , 
\label{eq:Evtheta}
\end{align} 
where $0<\theta<\pi$. From \eqref{eq:Apm} we then 
have $A_{\pm} = e^{\pm i \theta}$, 
and from \eqref{eq:free-m=1sol} it follows that the solution reads
\begin{align}
c_m = \sin(m\theta + \delta)
\ , 
\label{eq:free-cm-osc}
\end{align} 
where $\delta$ is 
determined in terms of $\alpha$ and $\theta$ from 
\begin{align}
\cot\delta = \frac{1}{\alpha \sin\theta}
- \cot\theta , 
\label{eq:free-cotdelta}
\end{align}
understood for $\alpha=0$ in the limiting sense $\delta=0$. 
Note that $\delta$ is unique up to an additive integer multiple of~$\pi$. 
Note also that in the special cases 
$\alpha=0$, $\alpha=1$ and $\alpha=-1$, 
we have respectively 
$\delta=0$, $\delta = (\pi-\theta)/2$ and $\delta = -\theta/2$. 
As the solutions are oscillatory at $m\to\infty$, they are 
lattice analogues of plane waves, belonging to 
the continuous spectrum. 

Suppose then that $E< 0$. We now have 
$A_+>1$ and $0<A_-<1$, and avoiding an exponential 
divergence at $m\to\infty$ in 
\eqref{eq:free-gensol} requires $a=0$. Matching to 
\eqref{eq:free-m=1sol} is possible iff $\alpha>1$, 
and $E$ is then uniquely determined in terms of $\alpha$ by 
\begin{align}
E = \bigl(2 - \alpha - \alpha^{-1}\bigr) \lscale^{-2}
\ .  
\label{eq:Evalpha}
\end{align}
The solution is the normalizable, proper eigenstate $c_m = \alpha^{-m}$. 

Suppose next that $E>4\lscale^{-2}$. A similar analysis shows 
that a solution exists iff $\alpha<-1$, 
$E$ is uniquely determined in terms of $\alpha$ by~\eqref{eq:Evalpha}, 
and the solution is the normalizable, proper eigenstate $c_m = \alpha^{-m}$. 

Suppose finally that $E=0$ or $E=4\lscale^{-2}$. 
As the solutions \eqref{eq:free-cm-exceptional} are not normalizable, 
these special values of $E$ are not proper eigenvalues, 
and as points of measure zero they do not contribute to the spectrum. 

We summarize. For any $\alpha$, the spectrum contains 
the band $(0,4\lscale^{-2})$ of improper, Dirac-delta-normalizable
eigenstates, given by \eqref{eq:free-cm-osc} with 
\eqref{eq:Evtheta} and~\eqref{eq:free-cotdelta}. 
For each $\alpha$ such that $|\alpha|>1$, 
there exists in addition a unique proper  
eigenstate, given by $c_m = \alpha^{-m}$ 
and having the eigenvalue~\eqref{eq:Evalpha}. 
For $\alpha>1$ the proper eigenstate is a ground state at 
$E<0$, below the continuous band, 
and for $\alpha<-1$ it is a highly excited state at 
$E>4\lscale^{-2}$, above the continuous band. 

We note that the special case $\alpha=1$ of our free particle
Hamiltonian is obtained from the family of discrete Hamiltonians
(2.12) in \cite{odake-sasaki:review} with the choices $B(x)= D(x) =1$
for $x>0$ and $B(0)=1$, where the discrete configuration variable $x$
of \cite{odake-sasaki:review} is such that $x=0$ corresponds to our
$m=1$.  As discussed above, the spectrum with this value of $\alpha$
consists of the continuum band $(0,4\lscale^{-2})$ and does not
contain a bound state.  This exemplifies the observation made at the
end of subsection \ref{subsec:halfline} that even though we invoked a
fictitious $m=0$ lattice point to motivate the definition of the
Hamiltonian, the actual definition does not refer to this lattice
point and the value of the wave function at the fictitious lattice
point never enters the theory: in the formulation of
\cite{odake-sasaki:review} the $\alpha=1$ theory is indeed constructed
without a motivation via fictitious lattice points.

\subsubsection{Continuum limit}

We now show that the full one-parameter family of 
continuum Schr\"odinger quantizations 
can be recovered in the continuum limit. 

The key for taking the limit is to make 
$\alpha$ dependent on $\lscale$ in a suitable way. 
We introduce a parameter 
$L_P\in \BbbR \cup \{\infty\}$, 
which is considered 
independent of~$\lscale$, and 
we choose 
\begin{equation}
\alpha = 
\begin{cases}
0 & 
\text{for $L_P=0$;} 
\\[1ex]
1 & 
\text{for $L_P=\infty$;} 
\\[1ex]
{\displaystyle \frac{1}{1-(\lscale/L_P)}} 
& 
\text{otherwise,} 
\end{cases} 
\label{eq:free-alpha-LP}
\end{equation}
which is well defined for sufficiently 
small $\lscale$ when $L_P$ is in the interval $(0,\infty)$, 
and for all $\lscale$ when $L_P$ takes other values. 
When interpreted in terms of the 
fictitious lattice point at $m=0$ for which 
$c_0 = \alpha c_1$, 
\eqref{eq:free-alpha-LP} amounts to
\begin{equation}
\begin{aligned}
{\displaystyle c_0 + L_P \frac{(c_1 - c_0)}{\lscale} =0} 
&\phantom{x}
& 
\text{for $L_P\ne\infty$;} 
\\[1ex]
{\displaystyle \frac{(c_1 - c_0)}{\lscale} =0} 
&& 
\text{for $L_P=\infty$.} 
\end{aligned} 
\label{eq:free-LP-cond}
\end{equation}
Comparison of \eqref{Robin1} and \eqref{eq:free-LP-cond} suggests that 
the polymer theory should reduce to the continuum theory with $L =
L_P$ as $\lscale\to0$. 
We shall verify that it does. 

Let first $0 < L_P<\infty$ and focus on the ground state. 
By~\eqref{eq:Evalpha}, the ground state energy 
is given by $E = - L_P^{-2} +O(\lscale)$,
and the ground state wave function 
can be written in terms of the discrete 
distance coordinate $x=m\lscale$ as 
\begin{align}
\psi(x) & := c_{x/\lscale} = 
\left(1-\frac{\lscale}{L_P}\right)^{x/\lscale}
= e^{-x/L_P} +O(\lscale), 
\end{align} 
where in the last expression $L_P$ and $x$ are considered 
independent of~$\lscale$. In 
the limit $\lscale\to0$, we therefore recover the 
ground state energy and the ground state wave function 
of the continuum theory with $L =L_P$. 

Let then $L_P$ be arbitrary  
and consider energies in the continuous band. 
For fixed~$E$, 
\eqref{eq:Evtheta} and
\eqref{eq:free-cotdelta}
give 
\begin{equation}
\delta = 
\begin{cases}
0 & 
\text{for $L_P=0$;} 
\\
\tfrac12 \pi +  O(\lscale) & 
\text{for $L_P=\infty$;} 
\\
{\displaystyle - \arctan\Bigl(L_P \sqrt{E}\,\Bigr) +  O(\lscale)} 
& 
\text{otherwise,} 
\end{cases} 
\end{equation}
and
\begin{align}
\psi(x) & := c_{x/\lscale} = \sin\bigl((\theta/\lscale)x +
\delta\bigr)
= \sin\bigl(\sqrt{E}\,x + \delta\bigr) + O(\lscale^3) . 
\end{align} 
In the limit $\lscale\to0$, we therefore recover the continuum theory 
wave functions with $L =L_P$. 

We emphasize that the fine-tuning of $\alpha$ as a function of
$\lscale$ in \eqref{eq:free-alpha-LP} is essential.  If the limit
$\lscale\to0$ is instead taken with fixed~$\alpha$, the choice
$\alpha=1$ (corresponding to $L_P=\infty$ in~\eqref{eq:free-alpha-LP})
gives the Neumann continuum theory, $L=\infty$, while any other choice
for $\alpha$ (corresponding to $L_P\to0$ in~\eqref{eq:free-alpha-LP})
gives the Dirichlet continuum theory, $L=0$.

\section{Coulomb potential}
\label{sec:hydrogen}

As the next example, we consider 
the attractive Coulomb potential, 
$V(x) = - 1/x$. 
A~new feature is that this potential 
is singular at $x=0$. 

\subsection{Schr\"odinger half-line}

In Schr\"odinger quantization~\cite{fewster}, the
Hamiltonian has a one-parameter family of self-adjoint extensions,
specified by a boundary condition at $x=0$.  The boundary condition
does not take the Robin form \eqref{Robin1} but reads instead
\begin{align}
\bigl(\Lpar - \Psi(1) - \Psi(2) \bigr) \, \psi(0) 
+ \lim_{x\to0} \left( \frac{\psi(x) - \psi(0)}{x} + \psi(0) \ln x \right) 
=0 , 
\label{eq:hyd-cont-bc}
\end{align}
where $\psi$ is the wave function, 
$\Psi$ is the digamma function \cite{nist-dig-library} 
and $\Lpar \in \BbbR \cup \{\infty\}$ is the parameter. 
The second term in \eqref{eq:hyd-cont-bc} can be understood as a 
logarithm-corrected first derivative, 
where the correction must be included because 
of the singularity of the potential at $x\to0$. 

The spectrum consists of the positive continuum and a 
countable number of proper negative eigenvalues. 
Writing $s := 1/\sqrt{-4E}$ for $E<0$, 
the eigenvalues are given by the solutions to 
\begin{align}
\Lpar = \Psi(1-s) - \ln s + \frac{1}{2s} 
\ . 
\label{eq:hyd-s-eq}
\end{align}

The solutions to \eqref{eq:hyd-s-eq} with $\Lpar = \infty$ are $s =
1,2,3,\ldots$, coming from the simple poles of $\Psi$ at non-positive
integers. This extension is analogous to the Dirichlet boundary
condition for regular potentials, and it arises via a limiting
prescription from regularized potentials that make the
three-dimensional attractive Coulomb Hamiltonian essentially
self-adjoint~\cite{fewster}. This is the extension that yields the
textbook spectrum, realized in nature.

We note in passing that if physical dimensions are restored, 
$\Lpar$ has an interpretation as an inverse length scale.
Using $\Lpar$, rather than the associated 
length scale $1/\Lpar$~\cite{fewster}, 
is convenient for us because the spectrum changes 
smoothly across $\Lpar=0$. 
The only discontinuous 
change in the spectrum in terms of $\Lpar$ 
occurs when $\Lpar\to\infty$ through positive values: 
in this limit the ground state descends to $-\infty$ and disappears.

\subsection{Polymer half-line}

The half-line polymer Hamiltonian is given by 
$\widehat{H}_\alpha$ \eqref{eq:fHm-action} 
with $V(m\lscale)=-1/(m\lscale)$. 
As noted in subsection~\ref{subsec:halfline}, 
the boundedness of $V$ over the lattice points
shows that $\widehat{H}_\alpha$ is essentially self-adjoint. 

We focus on a numerical study of the proper  
eigenvalues in the limit $\lscale\to0$.  
The numerical scheme is as described in \cite{coulomb}
except that the boundary condition to shoot for is 
now $c_0 = \alpha c_1$. 

The key is again to make $\alpha$ depend on 
$\lscale$ in a suitable way. Motivated by 
the continuum boundary condition~\eqref{eq:hyd-cont-bc}, 
we introduce a parameter 
$\LparP \in \BbbR \cup
\{\infty\}$, 
and we choose 
\begin{equation}
\alpha = 
\begin{cases}
0 & 
\text{for $\LparP=\infty$;} 
\\[1ex]
{\displaystyle \frac{1}{1 - 
\lscale \bigl(\LparP - \Psi(1) - \Psi(2) + \ln \lscale \bigr)}} 
& 
\text{otherwise,} 
\end{cases} 
\label{eq:hyd-alpha-LtildeP}
\end{equation}
which is well defined for sufficiently small $\lscale$  
with any fixed value of~$\LparP$. 
Numerical results for the lowest three eigenvalues in terms of $\LparP$ 
are shown in Table~\ref{tab:hydrogen}. 
Numerical accuracy allows us to probe values of $\lscale$ 
down to~$10^{-5}$.

\begin{table}[p]
\scriptsize
\centering
\begin{tabular}{l|l l l l l l}
\multicolumn{7}{l}{$s_0$} \\[1ex]
\hline\hline
${{\LparP}}^{\vphantom{A^{\textstyle R}}}$ or $\Lpar$ 
                  & $10$        & $4/5$        & $0$         & $-1$        & $-10$       & $\infty$ \\
\hline
$\lscale = 1$       & --          & 0.1091       & 0.5745      & 0.7983      & 0.9934      & $1 + 2.9 \times 10^{-2}$ \\
$\lscale = 10^{-1}$ & 0.0290      & 0.4475       & 0.5483      & 0.6549      & 0.9168      & $1 + 3.1 \times 10^{-4}$ \\
$\lscale = 10^{-2}$ & 0.0636      & 0.4317       & 0.5225      & 0.6250      & 0.9073      & $1 + 3.1 \times 10^{-6}$ \\
$\lscale = 10^{-3}$ & 0.0660      & 0.4279       & 0.5178      & 0.6202      & 0.9062      & $1 + 3.1 \times 10^{-8}$ \\
$\lscale = 10^{-4}$ & 0.066170755 & 0.4273310795 & 0.517072743 & 0.619510329 & 0.906040222 & $1 + 3.1 \times 10^{-10}$ \\
$\lscale = 10^{-5}$ & 0.066190491 & 0.4272479722 & 0.516979832 & 0.619423583 & 0.906025030 & $1 + 3.1 \times 10^{-12}$\\
\hline
Schr\"odinger   & 0.0631 & 0.3814 & 0.4696 & 0.5785 & 0.9022 & 1 \\
\hline 
${{\LparP}}^{\vphantom{A^{\textstyle R}}} - \Lpareff$ \\
$\lscale = 10^{-3}$ & 0.397491563 & 0.4287638124 & 0.4303204240 & 0.431360980 & 0.44340833 & -- \\
$\lscale = 10^{-4}$ & 0.419935473 & 0.4236415186 & 0.4237432260 & 0.423857416 & 0.42477493 & -- \\
$\lscale = 10^{-5}$ & 0.422522628 & 0.4228930728 & 0.4229032450 & 0.422914658 & 0.42300644 & -- \\
\hline\hline
\multicolumn{7}{l}{} \\[1ex]
\multicolumn{7}{l}{$s_1$} \\[1ex]
\hline\hline
${{\LparP}}^{\vphantom{A^{\textstyle R}}}$ or $\Lpar$ 
                  & $10$        & $4/5$        & $0$         & $-1$         & $-10$      & $\infty$ \\
\hline
$\lscale = 1$       & 1.06769     & 1.436        & 1.645       & 1.806        & 1.9804     & $2 + 1.5 \times 10^{-2}$ \\
$\lscale = 10^{-1}$ & 1.09386     & 1.482        & 1.571       & 1.668        & 1.9172     & $2 + 1.6 \times 10^{-4}$ \\
$\lscale = 10^{-2}$ & 1.09981     & 1.463        & 1.545       & 1.638        & 1.9079     & $2 + 1.6 \times 10^{-6}$ \\
$\lscale = 10^{-3}$ & 1.10024     & 1.459        & 1.540       & 1.634        & 1.9068     & $2 + 1.6 \times 10^{-8}$ \\
$\lscale = 10^{-4}$ & 1.100263034 & 1.45830016   & 1.539302314 & 1.632903508  & 1.90664102 & $2 + 1.6 \times 10^{-10}$ \\
$\lscale = 10^{-5}$ & 1.100263299 & 1.45821961   & 1.539214395 & 1.632821310  & 1.90662595 & $2 + 1.6 \times 10^{-12}$ \\
\hline
Schr\"odinger     & 1.09630     & 1.417        & 1.496       & 1.595        & 1.9029     & 2 \\
\hline 
${{\LparP}}^{\vphantom{A^{\textstyle R}}} - \Lpareff$ \\
$\lscale = 10^{-3}$ & 0.420413025 & 0.4306943397 & 0.4307488453 & 0.436817400 & 0.44358639 & -- \\
$\lscale = 10^{-4}$ & 0.422779633 & 0.4237041296 & 0.4237847082 & 0.423885292 & 0.42478657 & -- \\
$\lscale = 10^{-5}$ & 0.422806855 & 0.4228993154 & 0.4229073956 & 0.422917448 & 0.42300773 & -- \\
\hline\hline
\multicolumn{7}{l}{} \\[1ex]
\multicolumn{7}{l}{$s_2$} \\[1ex]
\hline\hline
${{\LparP}}^{\vphantom{A^{\textstyle R}}}$ or $\Lpar$ 
                  & $10$        & $4/5$        & $0$          & $-1$        & $-10$ & $\infty$ \\
\hline
$\lscale = 1$       & 2.05400     & 2.4374       & 2.6463       & 2.8039      & 2.9755     & $3 + 1.033 \times 10^{-2}$ \\
$\lscale = 10^{-1}$ & 2.09419     & 2.4851       & 2.5734       & 2.6690      & 2.9172     & $3 + 1.042 \times 10^{-4}$ \\
$\lscale = 10^{-2}$ & 2.10026     & 2.4653       & 2.5467       & 2.6398      & 2.9080     & $3 + 1.042 \times 10^{-6}$ \\
$\lscale = 10^{-3}$ & 2.10069     & 2.4613       & 2.5421       & 2.6351      & 2.9069     & $3 + 1.042 \times 10^{-8}$ \\
$\lscale = 10^{-4}$ & 2.100712343 & 2.4607015522 & 2.54141508   & 2.63445759  & 2.90674777 & $3 + 1.042 \times 10^{-10}$ \\
$\lscale = 10^{-5}$ & 2.100712544 & 2.4606208222 & 2.54132730   & 2.63437572  & 2.90673263 & $3 + 1.042 \times 10^{-12}$ \\
\hline
Schr\"odinger 
                  & 2.09672     & 2.4193       & 2.4987       & 2.5970      & 2.9030     & 3 \\
\hline 
${{\LparP}}^{\vphantom{A^{\textstyle R}}} - \Lpareff$ \\
$\lscale = 10^{-3}$ & 0.420509240 & 0.4296974160 & 0.4306588703 & 0.431510579 & 0.44284175 & -- \\
$\lscale = 10^{-4}$ & 0.422787147 & 0.4237079042 & 0.4237880518 & 0.423888137 & 0.42479883 & -- \\
$\lscale = 10^{-5}$ & 0.422807636 & 0.4228997124 & 0.4229077629 & 0.422917713 & 0.42300750 & -- \\
\hline\hline
\end{tabular}
\caption{The table shows numerical results for
the lowest three eigenenergies in the Coulomb potential, 
in the polymer theory as a function of the 
parameter $\LparP$ and the scale~$\lscale$, 
and in the 
Schr\"odinger theory as a 
function of the parameter~$\Lpar$. 
The shown quantities 
$s_0$, $s_1$ and $s_2$ parametrize the eigenenergies $E_0$, $E_1$ and $E_2$
by $s = 1/\sqrt{-4E}$. 
The last three rows for each $s_i$ show the values of 
$\LparP- \Lpareff$ such that 
the Schr\"odinger eigenvalue with $\Lpar = \Lpareff$ 
equals the corresponding polymer eigenenergy. 
Note the convergence of $\LparP- \Lpareff$ to 
$0.423$ with decreasing~$\lscale$, 
within the numerical accuracy.}
\label{tab:hydrogen}
\end{table}

For $\LparP = \infty$, we find that the polymer eigenenergies converge to
those of the Schr\"odinger theory with $\Lpar = \infty$.  For
other values of~$\LparP$, 
we find that the polymer eigenenergies converge
to those of a Schr\"odinger theory with $\Lpar - \LparP \approx 0.423$.
While we do not have an analytic explanation for this shift in 
$\Lpar - \LparP$, we note that within our numerical accuracy the shift 
coincides with
the constant 
$\Psi(2) 
\approx 0.4227843351$ that
appears in \eqref{eq:hyd-cont-bc} and~\eqref{eq:hyd-alpha-LtildeP}. 
Examining whether this coincidence continues to hold beyond three decimal
places would require numerical work beyond $\lscale = 10^{-5}$.

If the limit $\lscale\to0$ is taken with any fixed value of~$\alpha$, 
the relation \eqref{eq:hyd-alpha-LtildeP} shows that 
the numerical results in Table \ref{tab:hydrogen} are consistent with 
convergence to the continuum theory with $\Lpar = \infty$.

\section{Scale invariant potential}
\label{sec:scaleinv}

We next consider the scale invariant potential, $V(x) = -\lambda/x^2$,
where $\lambda$ is a real-valued constant. Although we work in
dimensionless variables, we note that $\lambda$ remains dimensionless
even when physical dimensions are restored, and $\lambda$ 
is hence a pure number whose value is significant regardless any unit
choices. That the coupling constant is dimensionless is the special
property of a scale invariant potential.

\subsection{Schr\"odinger half-line}

Schr\"odinger quantization of the system is 
reviewed 
in~\cite{scale invariant}. For $\lambda \le -3/4$ the Hamiltonian is
essentially self-adjoint and the spectrum consists of the positive
continuum. For $\lambda > -3/4$ the self-adjoint extensions are
specified by a parameter that takes values in $\text{U}(1) \simeq
S^1$, and the spectrum of each extension contains the positive
continuum but discrete negative eigenenergies can also exist. We shall
recall here relevant facts about the cases in which discrete
eigenenergies do exist.

For $\lambda > 1/4$, each extension has 
a countable 
tower of proper eigenenergies, given by
\begin{align}
E_n
&= - \exp\Bigl[2(\gamma - \pi n)/\sqrt{\lambda - (1/4)} \, \Bigr] , 
\ \ n \in \mathbb{Z} ,
\label{eq:scaleinv-contEn}
\end{align}
where $\gamma \in [0,\pi)$ is the parameter specifying the extension.
The discrete spectrum is unbounded below, $E_n\to -\infty$ as
$n\to-\infty$, and it accumulates to $0$ from below, $E_n\to0_-$ as
$n\to\infty$.

For $\lambda = 1/4$, one of the extensions has no discrete
eigenenergies but the rest have exactly one negative eigenenergy each.

For $-3/4 < \lambda < 1/4$, there is an open interval of extensions
that have exactly one negative eigenenergy each. The remaining
extensions have no discrete eigenenergies.

\subsection{Polymer half-line}

The half-line polymer Hamiltonian is given by 
$\widehat{H}_\alpha$ \eqref{eq:fHm-action} 
with $V(m\lscale)=-\lambda/(m\lscale)^2$. 
The boundedness of $V$ over the lattice points again 
implies 
that $\widehat{H}_\alpha$ 
is essentially self-adjoint. 

We again focus on a numerical study of the discrete eigenvalues. 
Because of the scale invariance of the potential, the polymer
eigenvalues depend on $\lscale$ only through the overall
factor~$\lscale^{-2}$. The question to be examined hence is
how the combination $\lscale^2 E$ depends on~$\lambda$.

Because of the strong singularity of the continuum potential at
$x\to0$, it is not clear how the Schr\"odinger theory self-adjointness
boundary conditions \cite{scale invariant} might motivate a value of
$\alpha$ in our polymer Hamiltonian.  We hence take $\alpha =
\tan\chi$, where $\chi\in[0,\pi)$, allowing $\chi=\pi/2$ to be
considered as a limiting case. The numerical implementation is as in
\cite{scale invariant} except that the boundary condition to shoot for
is
\begin{align}
c_0 \cos\chi  - c_1 \sin\chi =0 . 
\end{align}

\begin{figure}[p]
\begin{center}
\includegraphics[scale=0.6]{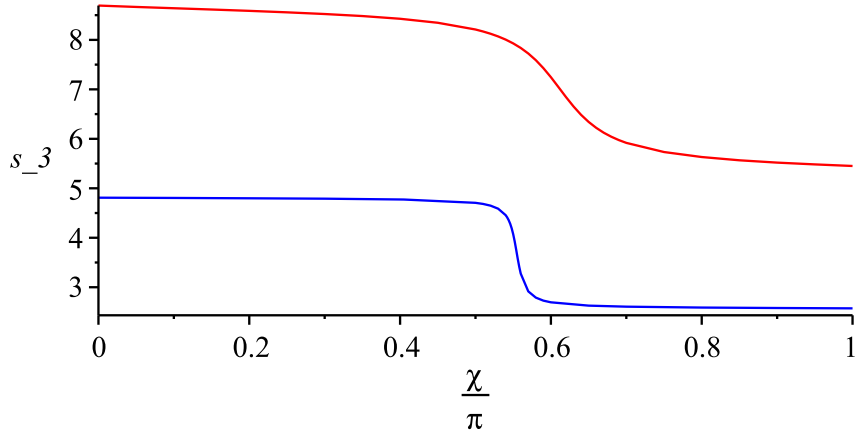}
\caption{Scale invariant potential. 
$s_3 := - \ln(-\lscale^2 E_3)$ is shown as a function of 
$\chi$ for $\lambda=4$ (upper curve, red) and 
for $\lambda=8$ (lower curve, blue).}  
\label{fig:scaleinvs3chi}
\end{center}
\end{figure}

For $\lambda \gtrsim2$, we find a tower of negative eigenenergies
$E_0< E_1 < E_2< \cdots<0$. We have followed the tower up to $E_6$ but
slowness of the numerics has not enabled us to examine whether the
tower terminates. Starting at $\chi=0$ and increasing~$\chi$, $E_0$
first migrates downwards and then disappears at $\chi\approx\pi/2$;
slowness of the numerics has not enabled us to examine whether this
disappearance happens by descent to $-\infty$ or by some other
mechanism. The excited states $E_n$, $n>0$, migrate down smoothly,
meeting at $\chi\to\pi$ with $E_{n-1}$ at $\chi=0$. Graphs for $E_3$
in terms of $\chi$ are shown in Figure \ref{fig:scaleinvs3chi} for
$\lambda=4$ and $\lambda=8$.

\begin{figure}[p]
\begin{center}
\includegraphics[scale=0.5]{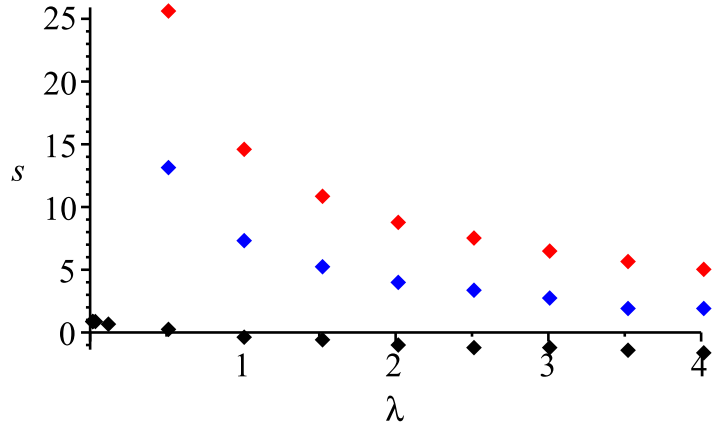}
\caption{Scale invariant potential. 
Values of $s_n := - \ln(-\lscale^2 E_n)$ 
are shown as a function of $\lambda$ for $\chi=\pi/3$, 
for 
$n=0$ (low, black), 
$n=1$ (middle, blue)
and 
$n=2$ (top, red). 
As $\lambda\to 0$, $s_0$ approaches the value of the $\lambda=0$ polymer theory, 
$-\ln(-2+4/\sqrt{3})\approx 1.11731$, and the excited states disappear.}  
\label{fig:scaleinvs0lambda}
\end{center}
\end{figure}

When $\lambda$ decreases towards~$0$, the numerics becomes
slow. Outside the interval $\pi/4 < \chi < \pi/2$ we have found no
evidence that any eigenstates would survive. Within the interval
$\pi/4 < \chi < \pi/2$, however, comparison with the free particle
polymer theory of Section \ref{sec:freeparticle} suggests that the
ground state should persist as $\lambda\to0$ and tend to that of the
corresponding $\lambda=0$ theory, given by
\begin{align}
E_0 = \frac{2}{\lscale^2} 
\left( -1 + \frac{1}{\sin2\chi} \right) , 
\end{align}
and we find numerical evidence that this indeed happens.
A~sample plot is 
shown in Figure~\ref{fig:scaleinvs0lambda}.

\begin{figure}[p]
\begin{center}
\includegraphics[scale=0.5]{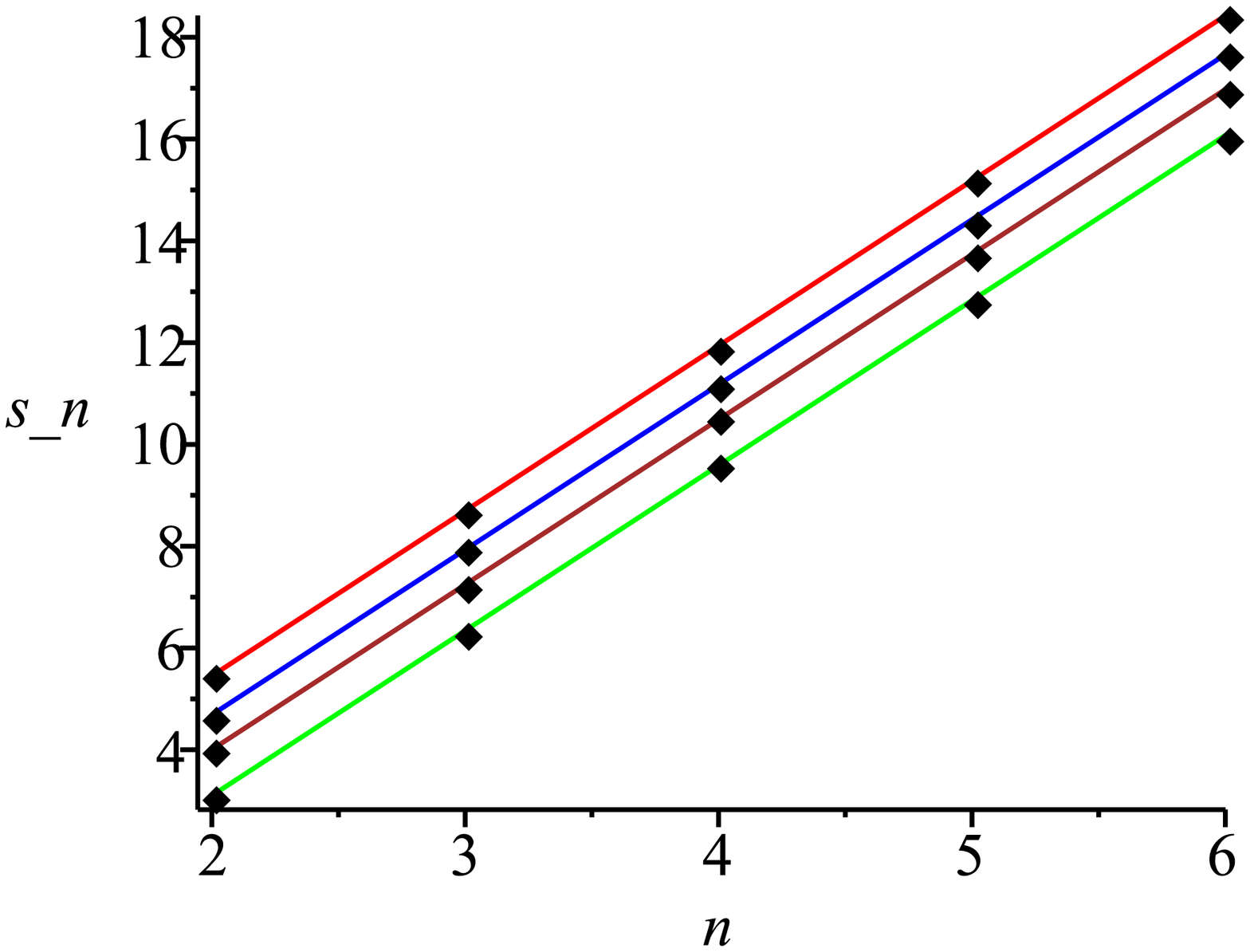}
\includegraphics[scale=0.5]{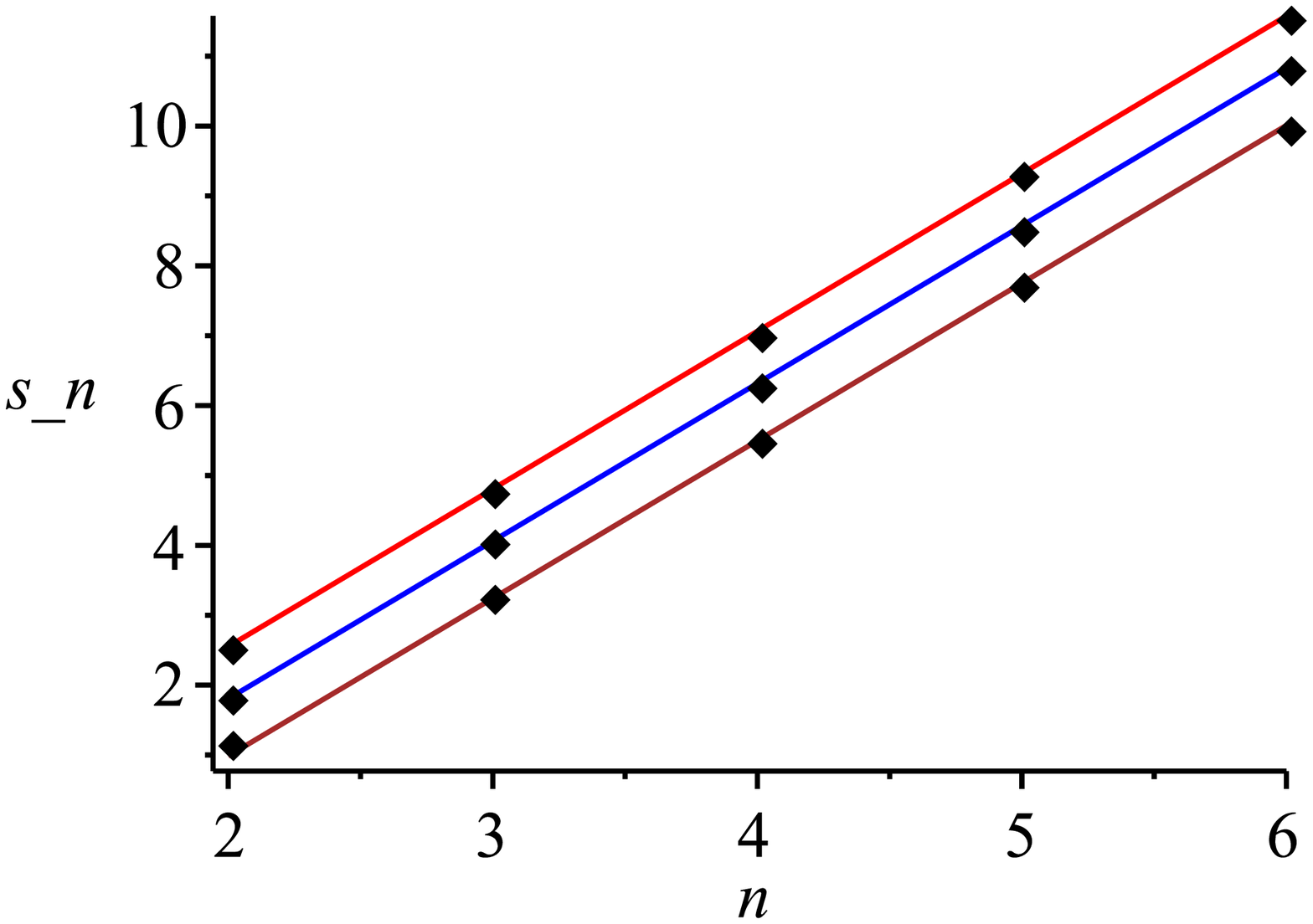}
\caption{Scale invariant potential. 
Values of $s_n := - \ln(-\lscale^2 E_n)$, 
$2\le n \le 6$, are shown for 
different values of~$\chi$, 
with a linear regression fit to the continuum 
formula~\eqref{eq:scaleinv-contEn}. 
Upper diagram shows 
$\lambda=4$
and the lines from top to bottom are with 
$\chi=0$ (red), 
$\chi=0.55\pi$ (blue), 
$\chi=0.60\pi$ (brown)
and
$\chi=0.65\pi$ (green). 
Lower diagram shows 
$\lambda=8$
and the lines from top to bottom are with 
$\chi=0$ (red), 
$\chi=0.55\pi$ (blue)
and 
$\chi=0.56\pi$ (brown).}
\label{fig:scaleinvslin}
\end{center}
\end{figure}

If the lowest few states are excluded, the eigenergies 
for $\lambda \gtrsim2$ are a good fit 
to the continuum eigenenergy formula~\eqref{eq:scaleinv-contEn}, as shown in 
Figure~\ref{fig:scaleinvslin}. 
The continuum parameter $\gamma$ is an increasing function of~$\chi$, 
and for large $\lambda$ it is slowly varying except when 
$\chi$ is
close to~$\pi/2$.

\section{Einstein-Rosen wormhole throat}
\label{sec:throat}

As the final example, we study the interior dynamics of the eternal
Schwarzschild black hole, adopting as the configuration variable
the area of the Einstein-Rosen wormhole throat and as the time
parameter the proper time of a comoving observer at the throat
\cite{friedman,redmount,Louko:1996md}.

\subsection{Schr\"odinger half-line}

Following~\cite{throata}, we take the classical phase space to be
$(\polv,\Pi)$, where the positive-valued configuration coordinate
$\polv$ is the square of the wormhole throat area-radius and $\Pi$ is
the real-valued conjugate momentum. The classical Hamiltonian is given
by 
\begin{align}
\label{eq:classH2}
H = \tfrac12 \, {\polv^{1/2}} 
\! \left( 4\Pi^2 + 1 \right) \,  . 
\end{align} 

We take the wave functions in Schr\"odinger quantization to be 
square integrable functions of $\polv$ in the measure~$d\polv$.
For reasons that will emerge in the polymer theory, 
we consider the quantum Hamiltonian 
with the symmetric ordering 
\begin{align}
\widehat{H}_\text{Schr} =
\frac{1}{2}\left(- 4 \frac{\partial}{\partial\polv} \polv^{1/2}
\frac{\partial}{\partial\polv}
+ \polv^{1/2}\right) . 
\label{eq:HS_phi}
\end{align}
It is shown in \cite{throata,Louko:1996md}
that the self-adjoint extensions 
of $\widehat{H}_\text{Schr}$
are labelled by the parameter $\theta \in [0,\pi)$ such that 
\begin{align}
\psi(\polv) =
\left[ 1 + O(\polv^{3/2}) \right] \sin\theta 
+ \polv^{1/2} \left[ 1 + O(\polv^{3/2}) \right] \cos\theta \, , 
\label{eq:throat-Sch-bc1}
\end{align} 
or equivalently 
\begin{align}
\psi(0) \cos\theta
- \left(\lim_{\polv\to0} 
\frac{\psi(\polv) - \psi(0)}{\polv^{1/2}}\right) \sin\theta =0 \, , 
\label{eq:throat-Sch-bc2}
\end{align}
so that $\theta=0$ is analogous to Dirichlet 
and $\theta=\pi/2$ is analogous to Neumann. 
The spectrum of each self-adjoint 
extension is discrete and bounded below. Further
analytic and numerical results on the spectrum are given
in~\cite{throata,Louko:1996md}.

\subsection{Polymer real line}

As the kinetic term of the classical Hamiltonian \eqref{eq:classH2} does
not have the conventional form of~\eqref{eq:class-Ham}, the polymer
quantization prescription of Section \ref{sec:polhalfline} must be
suitably generalized. We now recall the results of \cite{throata} 
for the situation where $\polv$ is polymer quantized on the full real line. 

We polymerize the pair $(\polv,\Pi)$ as the pair $(x,p)$ in
subsection~\ref{subsec:realline}, letting $\polv$ take all real values, 
defining the basis states 
\begin{align}
\psi_{\polv_0}(\polv) = \left\{
	\begin{array}{ll}
	1, & \polv=\polv_0\\
	0, & \polv\neq \polv_0
	\end{array}
\right.
\end{align}
with the inner product
\begin{align}
(\psi_\polv , \psi_{\polv^\prime}) = \delta_{\polv,\polv^\prime},
\label{eq:bohr-ip-polv}
\end{align}
and the operators 
\begin{subequations}
\begin{align}
\bigl(\hpolv \psi\bigr)(\polv) &:= \polv \psi(\polv), 
\label{polvact}
\\
\bigl(\widehat{U}_\mu \psi\bigr)(\polv) &:= \psi(\polv+\mu) , 
\label{throatUact}
\\
\hpi &:= \frac{1}{2i\mu}(\widehat{U}_\mu -
\widehat{U}_\mu^\dagger), 
\end{align}
\end{subequations} 
where the positive constant $\mu$ is again the fundamental polymer
length scale. The polymer counterpart of the Schr\"odinger
Hamiltonian \eqref{eq:H_phi} is taken to be 
\begin{align}
\widehat{H} =
\frac{1}{2}\Bigl(4 \hpi \hpolv^{1/2} \hpi
+ \hpolv^{1/2}\Bigr) . 
\label{eq:H_phi}
\end{align}
where the operator $\hpolv^{1/2}$ is defined by 
\begin{align}
\bigl(\hpolv^{1/2} \psi\bigr)(\polv) := 
{|\polv|}^{1/2} \, \psi(\polv)
\, . 
\end{align}

Specialising to the superselection sector in which 
$\polv$ is confined to the lattice 
$\bigl\{m \lscale \mid m\in\BbbZ \bigr\}$ with $\lscale := 2\mu$, 
and writing $\psi = \sum_m c_m \psi_{m\lscale}$, 
the action of 
$\widehat{H}$ on the two-sided square-summable sequence 
$c := {(c_m)}_{m=-\infty}^\infty$ reads
\begin{align}
\bigl(\widehat{H} c\bigr)_m 
= 
\frac{2}{\lscale^{3/2}}
\Bigl[
& 
\Bigl( \big| m+{\textstyle\frac{1}{2}} \big|^{1/2} 
+ \big| m-{\textstyle \frac{1}{2}} \big|^{1/2} 
+ \tfrac14 \lscale^2  {|m|}^{1/2} \Bigr) 
\, c_m 
\notag
\\
& 
- {\big| m+{\tfrac{1}{2}} \big| }^{1/2}\, c_{m+1}
- {\big| m-{\tfrac{1}{2}} \big| }^{1/2}\, c_{m-1}
\Bigr]
\, .
\label{eq:fH-action2-throat} 
\end{align}
It follows that the polymer Hilbert space breaks further into the even
superselection sector, in which $c_m=c_{-m}$, and the odd
superselection sector, in which $c_m=-c_{-m}$.  In the limit
$\lscale\to0$, it is found numerically \cite{throata} that the even
(respectively odd) sector converges to the Schr\"odinger theory with
$\theta=\pi/2$ ($\theta=0$). This convergence is consistent with what
one would expect by inspection of the Schr\"odinger theory boundary
condition~\eqref{eq:throat-Sch-bc1}.

\begin{table}[p]
\footnotesize
\centering
\begin{tabular}{l|l l l l l}
\multicolumn{5}{l}{$E_0$} \\[1ex]
\hline\hline
$\theta_P$ or $\theta$ & $0$     & $\pi/4$ & $\pi/2$ & $3\pi/5$ & $4\pi/5$ \\
\hline
$\lscale = 1$            & 1.23093 & 1.11371 & 0.904625 & 0.592200 & -- \\
$\lscale = 10^{-1}$      & 1.18178 & 1.04988 & 0.785426 & 0.423878 & -- \\
$\lscale = 10^{-2}$      & 1.16602 & 1.03780 & 0.770035 & 0.388607 & -- \\
$\lscale = 10^{-3}$      & 1.16113 & 1.03478 & 0.768373 & 0.380709 & -- \\
$\lscale = 10^{-4}$      & 1.15960 & 1.03390 & 0.768203 & 0.378624 & -- \\
\hline
Schr\"odinger          & 1.15890 & 1       & 0.768184 & 0.535489 & $-3.14921$ \\
\hline
$\theta_{\text{eff}}/\pi$ 
                       & 0       & 0.19527 & $1/2$    & 0.636933 & -- \\
\hline\hline
\multicolumn{5}{l}{} \\[1ex]
\multicolumn{5}{l}{$E_1$} \\[1ex]
\hline\hline
$\theta_P$ or $\theta$ & $0$     & $\pi/4$ & $\pi/2$ & $3\pi/5$ & $4\pi/5$ \\
\hline
$\lscale = 1$            & 1.87343 & 1.80220 & 1.72038 & 1.65342  & 1.39014 \\
$\lscale = 10^{-1}$      & 1.84343 & 1.75446 & 1.63581 & 1.54923  & 1.32377 \\
$\lscale = 10^{-2}$      & 1.83184 & 1.74589 & 1.62456 & 1.53274  & 1.30281 \\
$\lscale = 10^{-3}$      & 1.82825 & 1.74396 & 1.62336 & 1.53023  & 1.29678 \\
$\lscale = 10^{-4}$      & 1.82713 & 1.74343 & 1.62324 & 1.52975  & 1.29496 \\
\hline
Schr\"odinger          & 1.82661 & 1.72401 & 1.62322 & 1.56006 & 1.34989 \\
\hline
$\theta_{\text{eff}}/\pi$ 
                       & 0       & 0.19526 & $1/2$   & 0.63692 & 0.84817 \\
\hline\hline
\multicolumn{5}{l}{} \\[1ex]
\multicolumn{5}{l}{$E_2$} \\[1ex]
\hline\hline
$\theta_P$ or $\theta$ & $0$     & $\pi/4$ & $\pi/2$ & $3\pi/5$ & $4\pi/5$ \\
\hline
$\lscale = 1$            & 2.34359 & 2.28992 & 2.23644 & 2.19574  & 2.00416  \\
$\lscale = 10^{-1}$      & 2.32411 & 2.25111 & 2.16639 & 2.11021  & 1.95368  \\
$\lscale = 10^{-2}$      & 2.31424 & 2.24373 & 2.15637 & 2.09634  & 1.93652  \\
$\lscale = 10^{-3}$      & 2.31117 & 2.24216 & 2.15530 & 2.09435  & 1.93177  \\
$\lscale = 10^{-4}$      & 2.31022 & 2.24174 & 2.15519 & 2.09401  & 1.93036  \\
\hline
Schr\"odinger          & 2.30978 & 2.22684 & 2.15518 & 2.11361  & 1.97162  \\ 
\hline
$\theta_{\text{eff}}/\pi$ 
                       & 0       & 0.19524 & $1/2$   & 0.63691  & 0.84817  \\
\hline\hline
\multicolumn{5}{l}{} \\[1ex]
\multicolumn{5}{l}{$E_3$} \\[1ex]
\hline\hline
$\theta_P$ or $\theta$ & $0$     & $\pi/4$ & $\pi/2$ & $3\pi/5$ & $4\pi/5$ \\
\hline
$\lscale = 1$            & 2.73280 & 2.68919 & 2.64913 & 2.61958  & 2.46137  \\
$\lscale = 10^{-1}$      & 2.72106 & 2.65720 & 2.58907 & 2.54603  & 2.42038  \\
$\lscale = 10^{-2}$      & 2.71224 & 2.65049 & 2.57977 & 2.53351  & 2.40498  \\
$\lscale = 10^{-3}$      & 2.70948 & 2.64911 & 2.57877 & 2.53177  & 2.40087  \\
$\lscale = 10^{-4}$      & 2.70862 & 2.64875 & 2.57867 & 2.53148  & 2.39961  \\
\hline
Schr\"odinger          & 2.70822 & 2.63625 & 2.57866 & 2.54653  & 2.43449 \\ 
\hline
$\theta_{\text{eff}}/\pi$ 
                       & 0       & 0.19526 & $1/2$   & 0.63691  & 0.84816 \\
\hline\hline
\end{tabular}
\caption{The table shows 
the four lowest eigenenergies for the Einstein-Rosen wormhole throat, 
in the polymer theory as a function of the parameter 
$\theta_P$ and the scale~$\lscale$, 
and in the Schr\"odinger theory as a function of the parameter~$\theta$. 
For each $\theta_P$, the polymer eigenenergies converge to those of the 
Schr\"odinger theory with $\theta = \theta_{\text{eff}}$ as shown.} 
\label{tab:nonsing-throat}
\end{table}

\subsection{Polymer half-line}

We wish to modify the 
polymer Hamiltonian \eqref{eq:fH-action2-throat} to act on 
\emph{one-sided\/} square-summable sequences, 
of the form $c := {(c_m)}_{m=1}^\infty$ and with the inner product
$\left( d, c \right) = 
\sum_{m=1}^\infty \overline{d_m} \, c_m$. 
We again choose the one-parameter family of modified Hamiltonians 
$\bigl\{\widehat{H}_\alpha \mid \alpha\in\BbbR\bigr\}$ 
to be defined by adding at $m=0$ a fictitious lattice point with  
with $c_0=\alpha c_1$, so that 
\begin{subequations}
\label{eq:fH-actionalpha-throat} 
\begin{align}
\bigl(\widehat{H} c\bigr)_m 
&= 
\frac{2}{\lscale^{3/2}}
\Bigl[
\Bigl( \big( m+{\textstyle\frac{1}{2}} \big)^{1/2} 
+ \big( m-{\textstyle \frac{1}{2}} \big)^{1/2} 
+ \tfrac14 \lscale^2  {m}^{1/2} \Bigr) 
\, c_m 
\notag 
\\
& 
\hspace{10ex}
- {\big( m+{\tfrac{1}{2}} \big) }^{1/2}\, c_{m+1}
- {\big( m-{\tfrac{1}{2}} \big) }^{1/2}\, c_{m-1}
\Bigr]
\ \ \ \text{for $m>1$}, 
\\[1ex]
\bigl(\widehat{H} c\bigr)_1
& = 
\frac{2}{\lscale^{3/2}}
\Bigl[
\Bigl( \big( {\textstyle\frac{3}{2}} \big)^{1/2} 
+ (1-\alpha) \big({\textstyle \frac{1}{2}} \big)^{1/2} 
+ \tfrac14 \lscale^2  \Bigr) 
\, c_1 
- {\big({\tfrac{3}{2}} \big) }^{1/2}\, c_{2}
\Bigr] \, . 
\end{align}
\end{subequations}
Each $\widehat{H}_\alpha$ is symmetric, 
and the coefficient of $\overline{d_1} \, c_1$ in the 
inner product cannot be changed
without losing this symmetricity. 
The growth of the potential term suggests, 
by comparison 
with 
the solutions to the continuum eigenvalue equation, 
that each $\widehat{H}_\alpha$ is essentially self-adjoint; 
we have however not attempted to examine this rigorously. 

We wish to examine the bound states in the limit $\lscale\to0$. 
As in the previous cases, the key is to make $\alpha$ depend on $\lscale$ in a suitable way. 
Motivated by the continuum boundary condition \eqref{eq:throat-Sch-bc2}, 
we introduce a parameter 
$\theta_P \in [0, \pi)$, 
and we choose 
\begin{align}
\alpha = \frac{\sin\theta_P}{\sqrt{\lscale} \cos\theta_P + \sin\theta_P}
\, , 
\label{eq:throat-alpha-thetaP}
\end{align}
which is well defined for sufficiently small $\lscale$ with any fixed value of~$\theta_P$. 
Numerical results for the four lowest eigenvalues in terms of $\theta_P$ 
are shown in Table~\ref{tab:nonsing-throat}. 

\begin{figure}[t!]
\begin{center}
\includegraphics[scale=0.5]{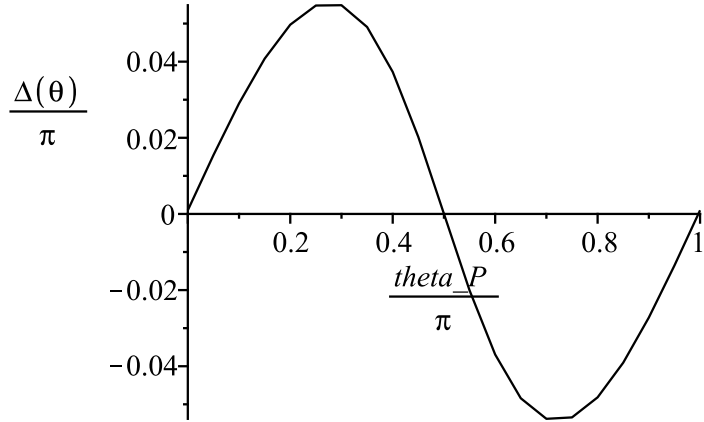}
\caption{$\Delta(\theta) := (\theta_P - \theta_{\text{eff}})$ 
is shown as a function of~$\theta_P$ 
for the Einstein-Rosen wormhole throat.} 
\label{fig:throatdeltatheta}
\end{center}
\end{figure}

For $\theta_P =0$ and $\theta_P =\pi/2$, 
the polymer eigenenergies converge to those 
of the Schr\"odinger theory with 
respectively $\theta =0$ and  $\theta =\pi/2$: 
this reproduces the results found in the 
even and odd sectors of the real line polymer theory in~\cite{throata}. 
For other values of $\theta_P$ the polymer 
eigenenergies converge to
those of the Schr\"odinger theory with $\theta=\theta_{\text{eff}}$, 
which differs from~$\theta_P$ as shown in 
Figure~\ref{fig:throatdeltatheta}. 
Note that 
when $\theta_P$ runs over the full interval $[0,\pi)$, 
so does $\theta_{\text{eff}}$. 

If the limit $\lscale\to0$ is taken with $\alpha=1$, 
it is seen from \eqref{eq:throat-alpha-thetaP} and 
Table \ref{tab:nonsing-throat} that the polymer theory converges to the 
continuum theory with $\theta=\pi/2$, 
and if the limit is taken with any other fixed value of~$\alpha$, 
the polymer theory converges to the continuum theory with $\theta=0$. 
The continuum Dirichlet-type boundary 
condition hence again emerges as generic.


\section{Conclusions}
\label{sec:conclusions}

We have constructed a one-parameter family of polymer quantization
Hamiltonians on the half-line. This family is mathematically analogous
to the one-parameter family of Robin boundary conditions in
Schr\"odinger quantization on the half-line.  For the free particle,
the attractive Coulomb potential and the Einstein-Rosen wormhole
throat, we found that the full family of continuum Robin boundary
conditions can be recovered in the continuum limit provided the
polymer parameter is suitably fine-tuned, while without such a
fine-tuning the continuum limit yields a Dirichlet-type continuum
boundary condition. For the scale invariant potential the spectrum
depends on the polymer scale only by an overall scaling, and a proper
notion of a continuum limit does hence not exist, but even in this
case the spectral properties revealed a close correspondence between
the polymer theory parameter and the continuum theory boundary
condition.

A notable feature of our half-line Hamiltonian is that it is well
defined even when the classical potential function is singular at the
origin, without any need to regularize the singularity.
Mathematically, this feature arises because even though a fictitious
lattice point at the origin was invoked to motivate the
definition of the Hamiltonian, the actual value of the wave function at
the origin never enters the theory, neither through the kinetic term
nor through the potential term in the Hamiltonian.  
Physically, this
feature is relevant to the issue of singularity resolution in polymer
quantum gravity \cite{Bojowald:2006da,Ashtekar:2011ni} and may also
have some impact in situations where polymer quantization is adopted
as a regulator of ultraviolet divergences in quantum field
theory~\cite{Hossain:2009vd}.

One qualitative difference between the one-parameter family of
continuum half-line Hamiltonians and our one-parameter family
$\bigl\{\widehat{H}_\alpha\bigr\}$ 
of polymer half-line Hamiltonians is that the
continuum parameter takes values in $\text{U}(1) \simeq S^1 \simeq
\BbbR \cup \{\infty\}$ but our polymer parameter $\alpha$ takes values
in~$\BbbR$. While the definition \eqref{eq:fHm-action} 
of $\widehat{H}_\alpha$
does as such not have a well-defined limit as 
$\alpha\to\pm\infty$, one could ask whether the
(generalized) eigenstates of $\widehat{H}_\alpha$ 
might have an $\alpha\to\pm\infty$ limit that is
sufficiently regular for the family $\bigl\{\widehat{H}_\alpha\bigr\}$
to be completed into an~$S^1$. 
Heuristically, thinking of $\widehat{H}_\alpha$ in
terms of the fictitious lattice point at $m=0$ such that $c_0 = \alpha
c_1$, the $\alpha \to\pm \infty$ limit should mean taking $c_1=0$,
which in turn should be equivalent to $\widehat{H}_0$ on a
lattice that is shifted to the right by one step. For the free
particle, we can verify from the analytic
solution of Section \ref{sec:freeparticle} that this heuristics is
correct: as $\alpha\to\pm\infty$, \eqref{eq:Evalpha} shows that the bound
state disappears, while \eqref{eq:free-cotdelta} shows that states in
the continuous spectrum have the limit $\delta\to-\theta$, which gives
exactly the expected shift in $m$ in~\eqref{eq:free-cm-osc}.  It would
be interesting to examine whether the heuristics holds also for
nonconstant potentials.

In all our examples, the parameter in the half-line polymer
Hamiltonian affects the qualitative properties of the spectrum in a
way that mimics closely the effects of the self-adjointness parameter
in the continuum half-line theory. For example, for the free particle
we saw that a single normalizable state can be made to appear, and the
energy of this state can be tuned to any value outside the continuous
spectrum. This is just as in the continuum theory, the only difference
being that the polymer continuous spectrum is an interval while the
continuum continuous spectrum is half-infinite. As another example, we
saw instances where adjusting the polymer parameter makes the lowest
energy eigenvalue disappear while the higher eigenvalues each migrate
downwards by one notch: similar behaviour occurs in the continuum
theory. We were able to give a fully analytic description of these
phenomena only for the free particle, but our numerical results
suggest that it would be interesting to pursue an analytic description
also for nonconstant potentials.

The results of our case studies are mutually compatible, and
compatible with the continuum Schr\"odinger theory, to an extent that
may exemplify a generic behaviour of the half-line polymer theory as a
function of the parameter in the Hamiltonian. To examine this further,
it would of course be of interest to extend the case case studies to
wider classes of classical potentials, controlling both the analytic
and numeric aspects.  For example, one might want to consider the
harmonic oscillator, for which the Schr\"odinger problem is readily
analytically solvable in terms of parabolic cylinder
functions~\cite{nist-dig-library}. We expect the half-line harmonic
oscillator polymer theory to be related to the Schr\"odinger theory in
a way qualitatively similar to what we found for the wormhole throat;
however, we have not been able to develop a sufficient numerical
control of the polymer theory eigenvalues to examine this question
quantitatively.

\section*{Acknowledgements}

We thank Hanno Sahlmann, 
John Spevacek and an anonymous referee for useful 
comments, and in particular Don Marolf for raising the issue of 
the $\alpha\to\pm\infty$ limit. 
This work was supported in part by the Natural Sciences 
and Engineering Research Council of Canada, by STFC (UK) 
and by the National 
Science Foundation under Grant No.\ NSF PHY11-25915. 
G.~K. is grateful to the University of Nottingham for its 
hospitality during the initiation and completion of this work.
J.~L. thanks the organizers of the 
``Bits, Branes, Black Holes'' 
programme for hospitality at the 
Kavli Institute for Theoretical Physics,
University of California at Santa Barbara.

\end{document}